\documentclass[aps,prd,twocolumn,superscriptaddress,floats,amsfonts,amsmath,amssymb]{revtex4}

\usepackage{graphicx,epsf}
\usepackage{color}
\usepackage[active]{srcltx}
\usepackage{bm}
\usepackage[]{latexsym}

\newcommand{\bea}{\begin{eqnarray}}\newcommand{\eea}{\end{eqnarray}}
\newcommand{\be}{\begin{eqnarray}}\newcommand{\ee}{\end{eqnarray}}
\newcommand{\ba}{\begin{array}}\newcommand{\ea}{\end{array}}
\newcommand{\bit}{\begin{itemize}}\newcommand{\eit}{\end{itemize}}
\newcommand{\ben}{\begin{enumerate}}\newcommand{\een}{\end{enumerate}}

\usepackage[naustrian, english]{babel}

\usepackage{float}

\usepackage{hyperref}
\hypersetup{pdfauthor={huchiayu},pdftitle={},
            pdfsubject={},pdfcreator={pdfLaTeX with hyperref (\today)}}

\begin{document}
\restylefloat{figure}

\title{Simulation of the Event Reconstruction of Ultra High Energy Cosmic Neutrinos with Askaryan Radio Array}

\author{Shang-Yu Sun}
\email{r96244010@ntu.edu.tw}
\affiliation{Graduate Institute of Astrophysics, National Taiwan University, Taipei, Taiwan 10617}
\affiliation{Leung Center for Cosmology and Particle Astrophysics, National Taiwan University, Taipei, Taiwan 10617}
\author{Pisin Chen}
\email{chen@slac.stanford.edu}
\affiliation{Leung Center for Cosmology and Particle Astrophysics, National Taiwan University, Taipei, Taiwan 10617}
\affiliation{Department of Physics, National Taiwan University, Taipei, Taiwan 10617}
\affiliation{Graduate Institute of Astrophysics, National Taiwan University, Taipei, Taiwan 10617}
\affiliation{Kavli Institute for Particle Astrophysics and Cosmology, SLAC National Accelerator Laboratory, Menlo Park, CA 94025, U.S.A.}

\author{Melin Huang} 
\email{phmelin@snolab.ca}
\affiliation{Leung Center for Cosmology and Particle Astrophysics, National Taiwan University, Taipei, Taiwan 10617}
\begin{abstract}
Askaryan Radio Array (ARA), a large-scale radio Cherenkov observatory which scientists propose to develop in Antarctica, aims at discovering the origin and evolution of the cosmic accelerators that produce the highest energy cosmic rays by means of observing the ultra high energy (UHE) cosmogenic neutrinos. To optimize ARA's angular resolution of the incoming UHE neutrinos, which is essential for pointing pack to its source, the relation between the reconstruction capabilities of ARA and its design is studied. It is found that with the noise effect taken into account, in order to make this neutrino angular resolution as good as possible and detection efficiency as high as possible, the optimal choice for ARA geometry would be the station spacing of 1.6 km and the antenna spacing of 40 m.
\end{abstract}

\maketitle

%%%%%%%%%%%%%%%%%%%%%%%%%%%%%%%%%%%%%%%%%%%%%%%%%%%%%%%%%%%%%%%%%%%
\section{Introduction}
\begin{figure}[H]
	\begin{center}
	\includegraphics[width=8cm]{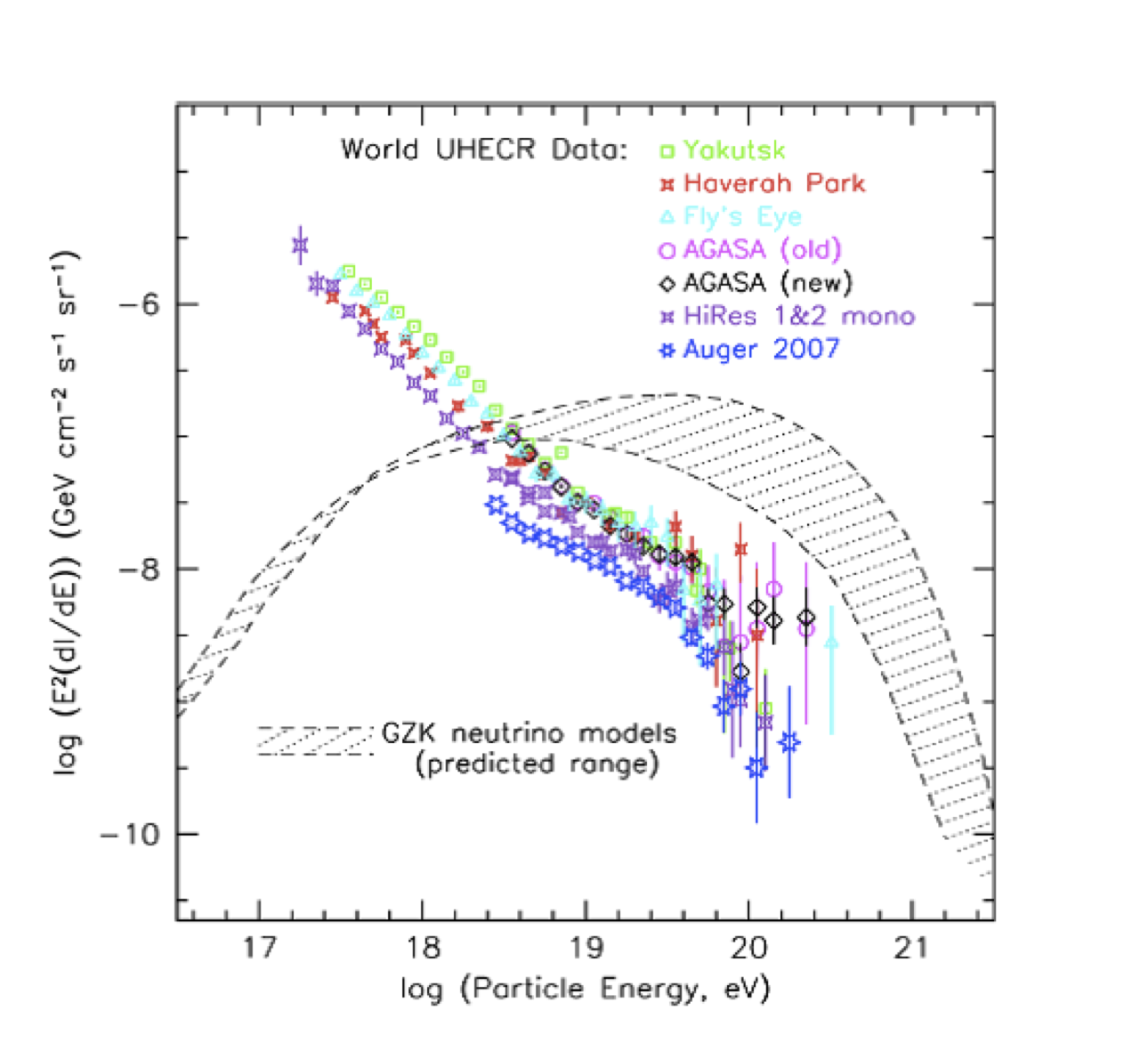}
	\caption{Energy spectrum of UHECR observation and GZK neutrino prediction.}
	\label{fig:nuspec}
	\end{center}
\end{figure}

%%%%%%%%%%%%%%%%%%%%%%%%%%%%%%%%%%%%%%%%%%%%%%%%%%%%%%%%%%%%%%%%%%%
     A limit on the cosmic ray energy was suggested in 1966 by Kenneth Greisen (US) \cite{ara5} and Vadim Kuzmin and Georgiy Zatsepin (Russia) \cite{ara6} independently based on interactions between the cosmic ray and the photons of the cosmic microwave background radiation. They predicted that cosmic rays with energies over the threshold energy of $6\times10^{19}$ eV would interact with cosmic microwave background photons to produce pions. This interaction would continue until their energies fall below the pion production threshold
This theoretical upper limit on the energy of cosmic rays from distant sources will create a cutoff in the cosmic ray spectrum right at the energy level of $6\times 10^{19}$ eV. And thus we call this GZK limit or GZK cutoff. Furthermore, the interaction of photons and protons does not stop at pion productions. These pions continue to decay into neutrinos. The whole interaction is named as GZK process, and the neutrinos produced from GZK process are called GZK neutrinos. The energy of GZK neutrino is also very high, still above the level of $10^{19}$ eV, so it can deserve the name of UHE neutrino.  Figure \ref{fig:nuspec} shows the energy spectrum of UHECR observation and GZK neutrino prediction. The GZK neutrino models in this figure were propsed by Kalashev \cite{ap19}, Protheroe, and Johnson \cite{ap18} et al., and UHECR observation data are taken from Auger \cite{ap15}, Yakutsk \cite{ap11}, the Fly's Eye \cite{ap16}, AGASA \cite{ap13}, HiRes \cite{ap14}, and Haverah Park \cite{ap12}. Error bars here only include statistical errors.

     Since UHE neutrinos can be a proof of the GZK process, but also a key to unveil the mystery of the cosmic accelerator, detection of them is scientifically important. Neutrinos cannot be directly observed, they but can be indirectly observed through their interactions with ordinary matter. There are three possible neutrino interactions: the elastic scattering, the charge current (CC) interaction, and the neutral current (NC) interaction. The outgoing high energy particles would, due to the Askaryan effect \cite{nsc2}, result in 20$\%$ excess of fast moving negative charges at the shower maximum and produce Cherenkov radiation, the radio band of which is coherent in ice and can be employed as a probe for UHE neutrinos. Askaryan effect states that a high energy particle which travels faster than the light in dense dielectric material can lead to charge asymmetry in its shower because positrons in the shower have higher probability of being stopped by the atoms in the medium and in addition the bound-state electrons tend to be ionized and comove with the shower. Then these fast moving access charges would cause Cherenkov radiation, where the radio and microwave bands of which are coherent. By invoking the Askaryan effect as a mechnism for detection, UHE neutrinos could in principle be observed. One of such attempts is the Antarctic Impulse Transient Antenna (ANITA) project, which intends to detect cosmogentic neutrinos interacting with the Antarctic ice. The ANITA Collaboration has performed an experiment at the Stanford Linear Accelerator Center (SLAC) in June 2006, which confirmed this effect in ice \cite{askice}.    
  
  Askaryan Radio Array (ARA), a large-scale radio Cherenkov observatory which scientists propose to develop in Antarctica, aims at discovering the origin and evolution of the cosmic accelerators that produce the highest energy cosmic rays by means of observing the ultra high energy (UHE) cosmogenic neutrinos. The reasons why ARA choose the Antarctic as the experiment site are the following: 1. There is plenty of ice as the target for detecting neutrinos. 2. The ice is so transparent to the RF shower signal that the spacing of RF detectors can be sufficiently large to enhance the effective volume and event rate. 3. It is more radio-quiet than other places in the world so as to reduce artificial signals considerably. 4. The temperature is so low that the background noise also reduces considerably.

     For these reasons, Antarctica is a very proper site to do the UHE neutrino experiment. With such a nature-given experimental environment, the next issue would be how to optimize the array geometry so as to maximize the performance. The primary goal of this paper work is to assess and optimize the capability of ARA, particularly the capability of reconstructing neutrino incoming directions, by means of Monte Carlo simulations.

\section{Simulation Method}\label{simu}
\subsection{Setting Array Geometry}

\begin{figure}
	\begin{center}
	\includegraphics[width=8cm]{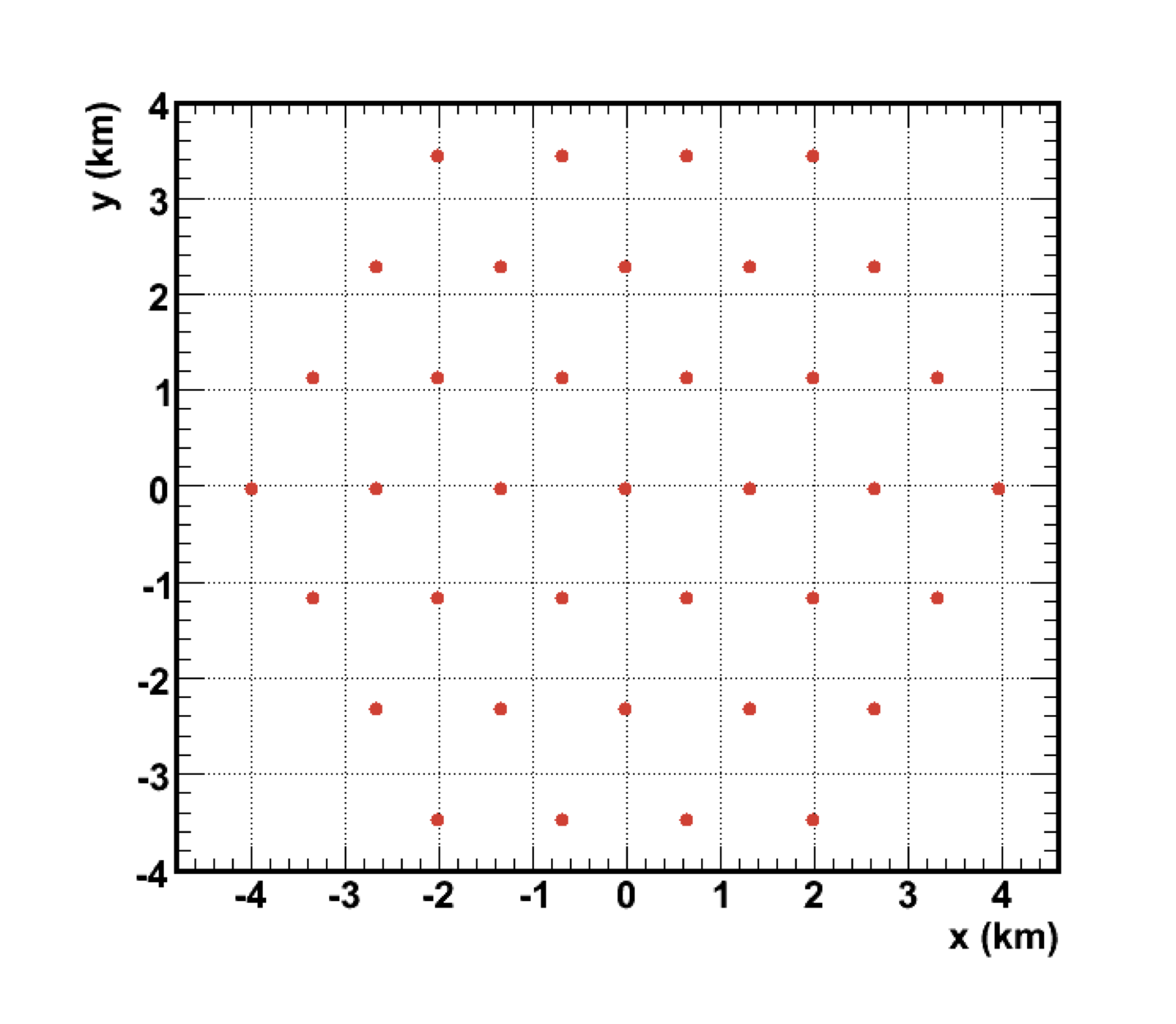}
	\caption{The geometry of the distribution and their coordinates.}
	\label{fig:ara}
	\end{center}
\end{figure}

     The proposed radio-based neutrino detector array, ARA, will eventually cover about 80 km$^2$ at the South Pole. There will be 37 antenna stations in the complete ARA. These 37 stations are located on a hexagonal lattice, as shown in Fig. \ref{fig:ara}, with a station spacing of 1.33 km. Note that the coordinate in this figure and in this analysis has its origin defined at the center of ARA, on the surface of the ice, and the z axis points to the sky. 
     
\begin{figure}
	\begin{center}
	\includegraphics[width=8cm]{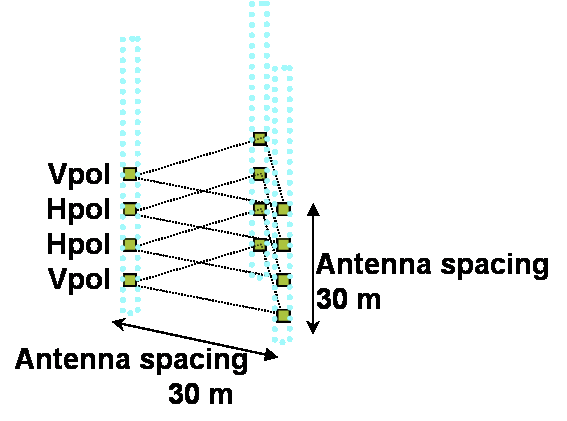}
	\caption{ARA antenna cluster geometry in a station, where there are twelve antennas, indicated by green squares.}
	\label{fig:station}
	\end{center}
\end{figure}

     Each station is composed of a triad of boreholes with depths of 200 m, on the corners of an equilateral triangle. Each borehole has four antennas, two of which are the horizontal-polarization (Hpol) antennas and the other two the vertical-polarization (Vpol) antennas, as shown in Fig. \ref{fig:station}. A pair of antennas, a Hpol one and a Vpol one, can detect the strengths of electrical field projected to horizontal plane (2D) and vertical line (1D) respectively, and the find the possible direction of the electric field. The side length of the equilateral triangle and the distance between two Vpol antennas in a borehole, are set the same, at 30 m. The location coordinate of antenna $i$ is denoted as $\mathbf{x}_i^{ant}$.

\subsection{Event Production}

     In the simulation, the shower events resulted from the CC or NC interactions are generated in the ice. This event generation does not differentiate neutrinos or anti-neutrinos, and flavors are not considered, either. In each event, 6 parameters are generated, including the shower location, $x_{sh}^{gen}, y_{sh}^{gen}, z_{sh}^{gen}$, the moving direction of neutrino: $\theta_\nu^{gen} ,\phi_\nu^{gen}$, and the intensity of the radio Cherenkov radiation induced by the shower followed by this interaction, $V_0^{gen}$. 
      We treat the shower location the same as the neutrino interaction vertex because of the small shower size in ice. The generated shower locations are uniformly distributed over a cylinder volume, where the center of the cylinder volume is located at the center of ARA. This volume has an axis along the vertical direction passing through the center of ARA, and has a radius of 6 km and a height from $z$ = -2 km to $z$ = 0. The reason why we set the height of this event cylinder as 2 km is that the thickness of the  ice in Antarctic is approximately 2 km. The choice of 6 km for the radius is due to the following reason. For the events with shower locations far away from ARA detectors, its radio radiation can not reach ARA because of attenuation. Therefore, the farthest distance for the radio signal of the event to travel to the ARA center is approximately estimated as $1 \rm{km} + 1.33 \rm{km} \times 3 = 5 \rm{km}$. For safety reason, we set it as 6 km rather than 5 km. Black dots are the generated shower locations, distributed uniformly in this field. Red circles represent the 37 stations.      

     The moving directions of the neutrinos are also set uniformly distributed isotropically over 4 $\pi$ solid angle. Furthermore, the outcome of a recorded waveform has been converted into voltage from electric field through the readout electronics and thus we set the initial intensity of the Cherenkov radiation in terms of the voltage, $V_0^{gen}$, in the range of 0 to 5 V.  

     At this stage, we generate six parameters for each event: $x_{sh}^{gen}$, $y_{sh}^{gen}$, $z_{sh}^{gen}$, $\theta_\nu^{gen}$, $\phi_\nu^{gen}$, $V_0^{gen}$. The first three are also denoted as the shower location vector, $\mathbf{x}_{sh}^{gen}$. The next two can also be described by a unit vector, $\mathbf{p}_\nu^{gen}$. There are 300 simulation events generated in each simulation.

     And these six event parameters are to be determined through $\chi^2$ fit in the next few sections.
    
%%%%%%%%%%%%%%%%%%%%%%%%%%%%%%%%%%%%%%%%%%%%%%%%%%%%%%%%%%%%%%%%%%%%%%%%%%%                
\subsection{Radio Cherenkov Wave from the shower location to Antennas}

     The Cherenkov radiation is set as a point source radiation because the shower size is of the order of $\sim$m and the propagation length before being received is $\sim$km. The radiation wave front has a cone shape with the apex at the interaction location, with the axis along the neutrino moving direction, and the span angle of the RF wave from 55 degree to 57 degree. The voltage waveform of this radiation signal is set as a bipolar wave:

\begin{equation}
  y= x\cdot e^{-x^2/2\sigma^2},
\end{equation}
where $y$ is voltage and $x$ stands for time.

     As this wave propagates through the ice, the wave intensity changes as 
\begin{equation}\label{vvxx}
\begin{split}
& V_i^{real}(\mathbf{x}_{sh}^{gen},V_0,\mathbf{p}_\nu^{gen})=V_0^{gen} \cdot \frac{D_0}{\sqrt{(\mathbf{x}_{sh}^{gen}-\mathbf{x}_i^{ant}})^2}\\
& \quad \times e^{\sqrt{(\mathbf{x}_{sh}^{gen}-\mathbf{x}_i^{ant})^2}/L_{att}^{ice}}\cdot a\cdot e^{-(\theta_i^{gen} - 56^{\circ})^2/2\sigma^2}\\
& \quad \times \left\{ \begin{array}{rl}
 \rm{sin}\alpha_{\it i}^{\it gen} &\mbox{(\rm for Hpol antenna)} \\
  \rm{cos}\alpha_{\it i}^{\it gen}&\mbox{(\rm for Vpol antenna),}
       \end{array} \right.
\end{split}
\end{equation}
where $D_0$ is the distance from the shower location to the location where $V_0^{gen}$ is measured, 1 km, $\theta_i^{gen}$ is the separation angle between the vector $\mathbf{p}_\nu^{gen}$ and the vector $\mathbf{x}_i^{ant}-\mathbf{x}_{sh}^{gen}$, and $\alpha_i^{gen}$ is the separation angle between the direction vector of the antenna $i$, $(x,y,z)=(0,0,1)$ and the direction vector of the electric field, $(\mathbf{x}_i^{ant}-\mathbf{x}_{sh}^{gen}) \times \Bigl[(\mathbf{x}_i^{ant}-\mathbf{x}_{sh}^{gen}) \times \mathbf{p}_\nu^{gen} \Bigr]$.

The travel time of this signal is
\begin{equation}
	t_i^{real}=\frac{\sqrt{(\mathbf{x}_{sh}^{gen}-\mathbf{x}_i^{ant}})^2}{c}, 
	\label{eq:t}
\end{equation}
where $c$ is the speed of light divided by the refraction index of ice.

     At the signal receiving end, the oscilloscope has time bin of 0.39 ns, and the time window is 100 ns. Noise before circuit has Gaussian distribution with mean voltage of 0 and $\sigma_{noise}$ 0.035 mV, whereas the trigger thresholds are that the Cherenkov cone intersects with the antenna and the attenuated signal must be larger than $7\sigma_{noise}$. An waveform is made by the following steps: an originally bipolar waveform magnified by a factor of signal strength $V_i^{real}$, shifted to the right by a time lag of $t_i$, and then added with noise.

%%%%%%%%%%%%%%%%%%%%%%%%%%%%%%%%%%%%%%%%%%%%%%%%%%%%%%%%%%%%%%%%%%%%%%%%%%%
\subsection{Determination of Arrival Time Difference and Pulse Voltage}

     To do the reconstruction of the events in the next stage, we have to extract arrival time difference, $\Delta t_i^{obs}$, and the pulse voltage, $V_i^{obs}$, from the waveform of each triggered antenna.

     The time when the signal arrives at the antennas should be precisely determined, and then with the difference of arrival time between any two antennas, and with the requirement that at least four antennas must be triggered, the shower location can be obtained through the process of fitting. One way can be applied to calculate arrival time, $t_i^{obs}$, for each antenna is the use of the point where V = 0 between the maximum and the minimum amplitudes.

     From the procedure described in the previous paragraph, for each antenna we can obtain an arrival time. Arrival time difference, which is the information actually used in the reconstruction, is the arrival time subtracted by the reference of the arrival time, $t_0^{obs}$, which is defined as the arrival time for the antenna receiving the strongest signal among all antennas. Therefore, 
\begin{equation}
	\Delta t_i^{obs} = t_i^{obs} - t_0^{obs}.
	\label{eq:deltat}
\end{equation}

     As for the pulse voltage, it is either the maximum point or the minimum point, depending on which one arrived at the antenna first.       
%%%%%%%%%%%%%%%%%%%%%%%%%%%%%%%%%%%%%%%%%%%%%%%%%%%%%%%%%%%%%%%%%%%%%%%%%%%
\subsection{Reconstruction of Neutrino Moving Directions}

     Our event reconstruction procedure is divided into two stages. The first stage is the reconstruction of shower location, $\mathbf{x}_{sh}^{gen}$. In this stage, the needed information is the arrival time difference, $\Delta t_i^{obs}$ for each antenna. We set up a $\chi^2$ formula: 
    
\begin{equation}
  \chi_1^2= \sum_{i} \frac{\bigl[\Delta t_i^{obs}-\Delta t_i^{hyp}(\mathbf{x}_{sh}^{hyp}) \bigr]^2}{\sigma_t^2}, 
	\label{eq:chi1}
\end{equation}
where $i$ is the index for all the triggered antenna, and $\Delta t_i^{hyp}$ is the hypothesized arrival time difference. By minimizing $\chi_1^2$, the best fit $\mathbf{x}_{sh}^{rec}$ can be found, where a grid search is employed. Local minima of $\chi_1^2$ value in the hypothesized-variable space is a serious problem and prohibit us from using other efficient ways to find the global minimum.

     In the second stage of reconstruction, we still use $\chi^2$ to find the best-fit. In this stage, the moving direction of neutrinos, $\mathbf{p}_\nu^{gen}$, is to be reconstructed, and the needed information is the pulse voltage received in each antenna. Furthermore, we also have to input the reconstructed shower location, $\mathbf{x}_{sh}^{rec}$, which is obtained in the first stage of reconstruction. Otherwise, we have to treat it as an unknown parameter to be reconstructed and this would intensively increase the computing time. The $\chi^2$ formula in the second stage is given as   
   
\begin{equation}
\begin{split}
&\chi_2^2= \\
&\sum_{i} \frac{\bigl[V_i^{obs}-V_i^{hyp}(\mathbf{x}_{sh}^{rec},V_0^{hyp},\mathbf{p}_\nu^{hyp})\bigr]^2 }{\sigma_V^2}, 
	\label{eq:chi2}
\end{split}
\end{equation}
where $i$ is the index for all the triggered antenna, and $V_i^{hyp}$ is the hypothesized pulse voltage. By minimizing $\chi_2^2$, the best fit $\mathbf{p}_\nu^{rec}=(1,\theta_\nu^{rec},\theta_\nu^{rec})$ can be found.
%\begin{equation}
%\begin{split}
%& \text{  ~  ~  ~  ~}V_i^{hyp}(\mathbf{x}_{sh}^{rec},V_0^{hyp},\mathbf{p}_\nu^{hyp})\\ 
%& \quad =\text{  }V_0^{hyp}\\
%& \quad \text{  ~ }\cdot \frac{D_0}{\sqrt{(\mathbf{x}_{sh}^{rec}-\mathbf{x}_i^{ant}})^2}\\
%& \quad \text{  ~ }\cdot e^{\sqrt{(\mathbf{x}_{sh}^{rec}-\mathbf{x}_i^{ant})^2}/L_{att}^{ice}}\\
%& \quad \text{  ~ }\cdot a\cdot e^{-(\theta_i^{hyp} - 56^{\circ})^2/2\sigma^2}\\
%& \quad \text{  ~ }\cdot \rm{sin}\alpha_i^{hyp} \text{   for Hpol antenna, or}\\ 
%& \quad \text{  ~  ~  }\rm{cos}\alpha_i^{hyp} \text{   for Vpol antenna}.\\
%\end{split}
%\label{eq:deltaVrec}
%\end{equation}

%%%%%%%%%%%%%%%%%%%%%%%%%%%%%%%%%%%%%%%%%%%%%%%%%%%%%%%%%%

%%%%%%%%%%%%%%%%%%%%%%%%%%%%%%%%%%%%%%%%%%%%%%%%%%%%%%%%%%%%%%%%%%%%%%%%%%%
\section{Results}\label{results}

\subsection{Resolutions of Shower Location, RF Wave Direction, and Neutrino Moving Direction}

\begin{figure}
	\begin{center}
	\includegraphics[width=8cm]{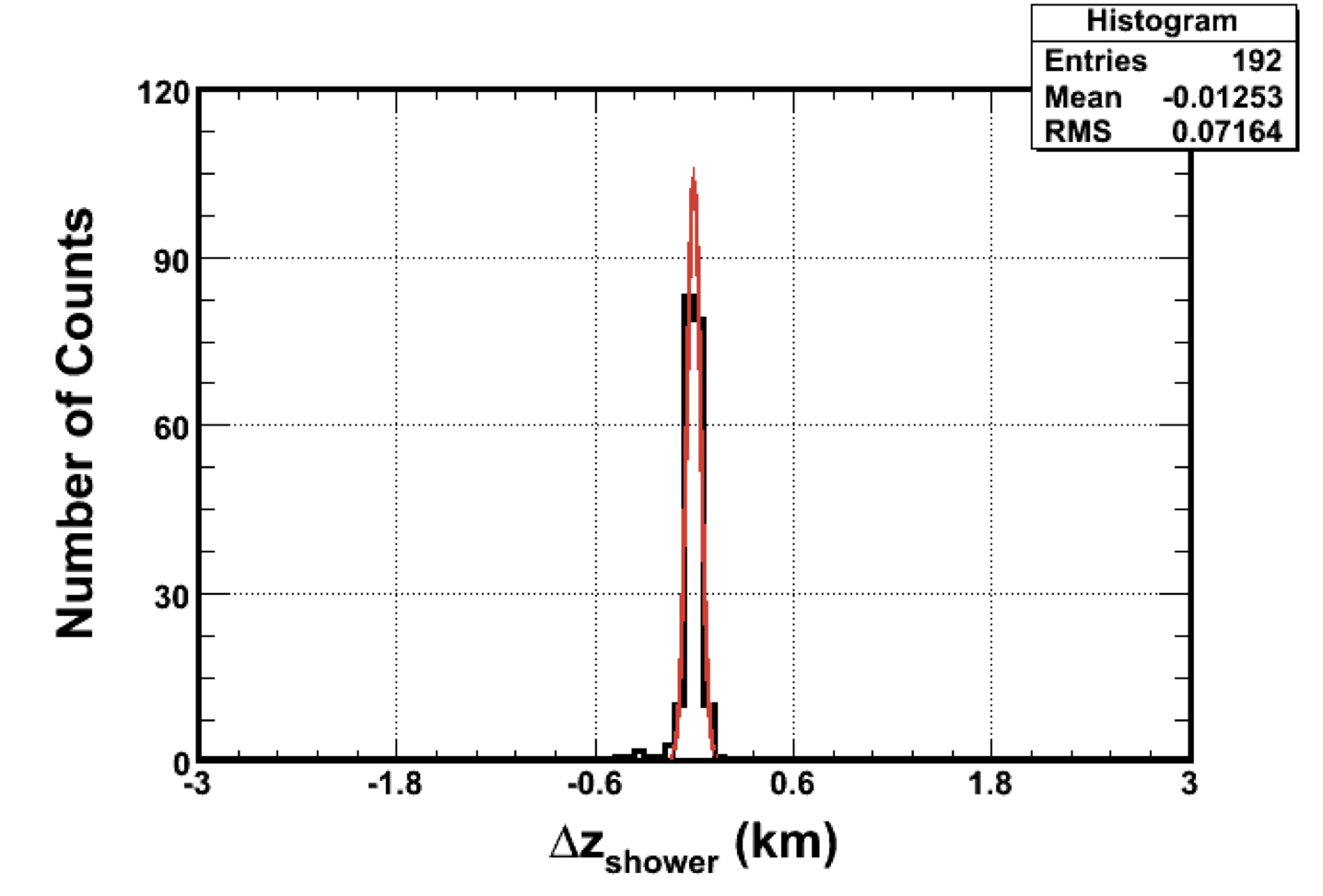}
	\caption{Resolution of shower location in z axis.}
	\label{fig:z}
	\end{center}
\end{figure}

     With Eq. \ref{eq:chi1}, the shower location can be reconstructed. The obtained resolutions of the shower location are 0.143 km in x axis, 0.098 km in y axis, 0.07 km in z axis. The resolution is the RMS value in Fig. \ref{fig:z}, which is the distributions of $\Delta z=z_{sh}^{rec}-z_{sh}^{gen}$. 
     
     Once the reconstructed shower locations are obtained, these reconstructed locations are taken as input in Eq. \ref{eq:chi2} for the reconstruction of neutrino moving directions. 
   
\begin{figure}
	\begin{center}
	\includegraphics[width=8cm]{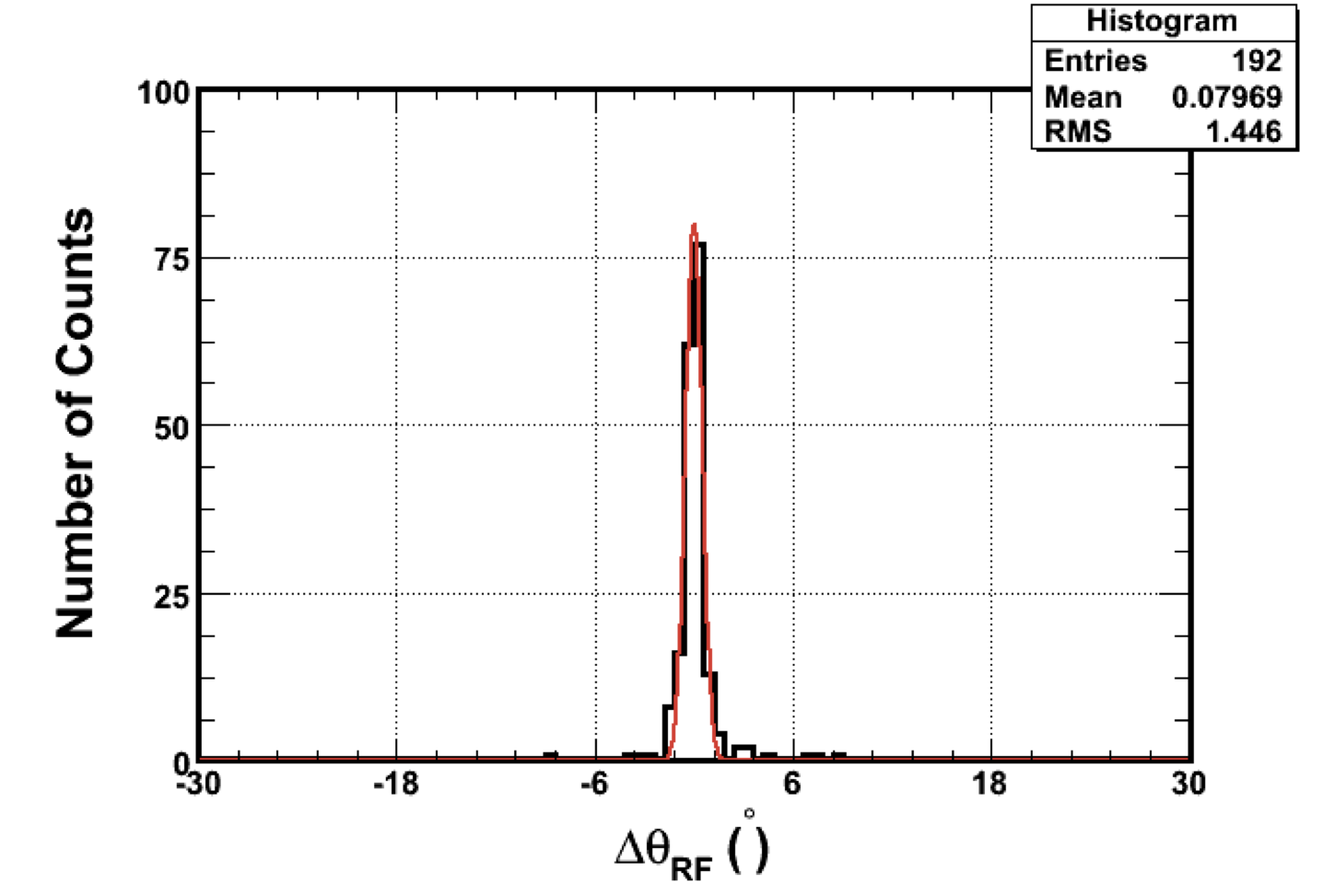}
	\caption{Resolution of RF wave direction in zenith angle.}
	\label{fig:rftheta}
	\end{center}
\end{figure}

     The obtained resolutions of the reconstructed RF wave direction, are 1.45$^\circ$ in $\theta$ direction , and 3.69$^\circ$ in $\phi$ direction. The resolution in $\theta$ direction is shown in Figs. \ref{fig:rftheta}, which is the distributions of $\Delta \theta_{RF}=\theta_{RF}^{rec}-\theta_{RF}^{gen}$. The $\theta$ and $\phi$ here are the zenith angle and the azimuthal angle of the spherical coordinate with origin defined as the location of the antenna receiving the strongest signal, and z axis as before. 

%%%%%%%%%%%%%%%%%%%%%%%%%%%%%%%%%%%%%%%%%%%%%%%%%%%%%%%%%%%%%%%%%%%%%%%%%%

\begin{figure}
	\begin{center}
	\includegraphics[width=8cm]{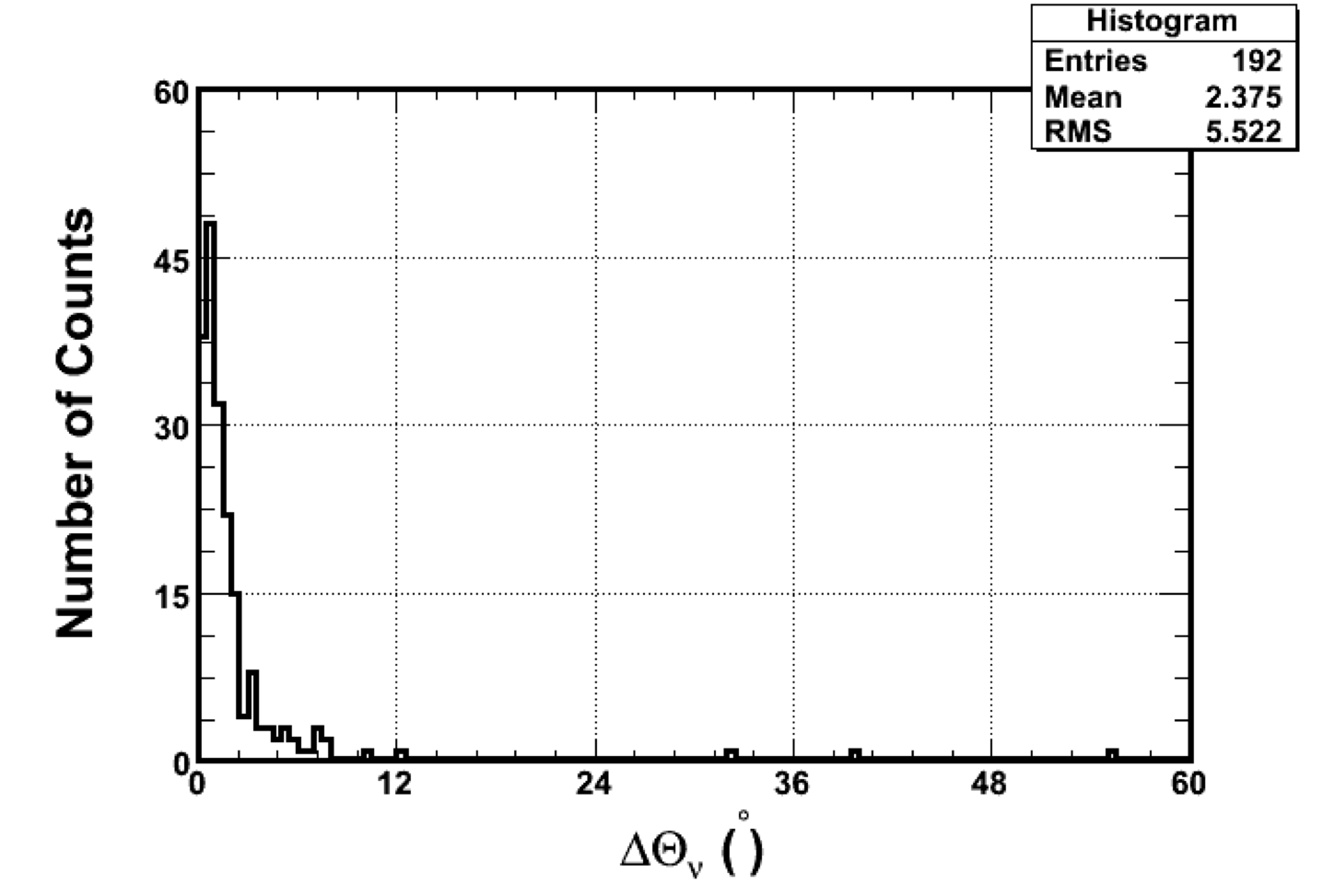}
	\caption{Distribution of the separation angle between the generated $\nu$ direction and the reconstructed one.}
	\label{fig:nuang}
	\end{center}
\end{figure}

     After the shower location and the RF wave direction are obtained, one can compute the neutrino direction according to Eq. \ref{eq:chi2}. The obtained resolutions, in $\theta$ direction 4.88$^\circ$, and in $\phi$ direction 3.76$^\circ$, which are the RMS values of the distributions of $\Delta \theta_{\nu}=\theta_{\nu}^{rec}-\theta_{\nu}^{gen}$, and $\Delta \phi_{\nu}=\phi_{\nu}^{rec}-\phi_{\nu}^{gen}$. The average of the separation angle between the generated $\nu$ direction and the reconstructed one is shown in Fig. \ref{fig:nuang}, which is 2.38$^\circ$.
     
     The average of this angle is taken for the comparison of the neutrino angular resolution in this analysis because the separation angles are always  positive, and thus the RMS value may not represent a proper indication of resolution. Note that $\theta$ and $\phi$ here are the zenith angle and the azimuthal angle of the spherical coordinate with the origin defined as the location of ARA center, and z axis as before. As mentioned before, the results presented so far have employed the ARA array geometry of station spacing as 1.33 km and antenna spacing as 30 m.

%%%%%%%%%%%%%%%%%%%%%%%%%%%%%%%%%%%%%%%%%%%%%%%%%%%%%%%%%%%%%%%%%%%%%%%%%%%
%\subsection{In Search of the Optimized Design: Angular Resolution of Neutrino Moving Direction and Detection Efficiency as Functions of Various Parameters}
\subsection{Optimization of ARA}
%Figure~\ref{fig:antspacing} is a graph showing how $\nu$ direction resolution behaves as antenna spacing changes from $10^{0.7}$ m to $10^{2.0}$ m with each step 0.1 in the power index of 10. All of the other parameters are fixed. The resolution is optimized when antenna spacing is between 30 m and 80 m. 

     To optimize the ARA, 16 different antenna spacings and 10 different station spacings are selected for the study on the resolution of the neutrino moving direction and the detection efficiency along with studies of noise effect.The optimum would be achieved when the resolution of the neutrino moving direction, i.e. $\langle\Delta \Theta_\nu\rangle$, is as good as possible, and the detection efficiency is as high as possible. The detection efficiency is defined as the number of triggered events that pass the trigger threshold divided by the total number of generated events in the cylinder volume, where the threshold applied to the pulse voltage is 7 $\sigma_{noise}$. 
  
     The antenna spacing varies from $10^{0.7}$ m to $10^{2.2}$ m in steps of 0.1 in the power index of 10. The station spacing changes from $1.33\rm{km}/5$ to $1.33\rm{km}\times 2$ in steps of $1.33\rm{km}/5$. Note that the antenna spacing means the distance from the top antenna to the bottom one. The vertical spacings between any two antennas are the same, and the center of the four antennas in a borehole is located at the depth of 200 m. In addition, the side of the equilateral triangle in a station is set the same as the antenna spacing.
     
\begin{figure}
	\begin{center}
	\includegraphics[width=8cm]{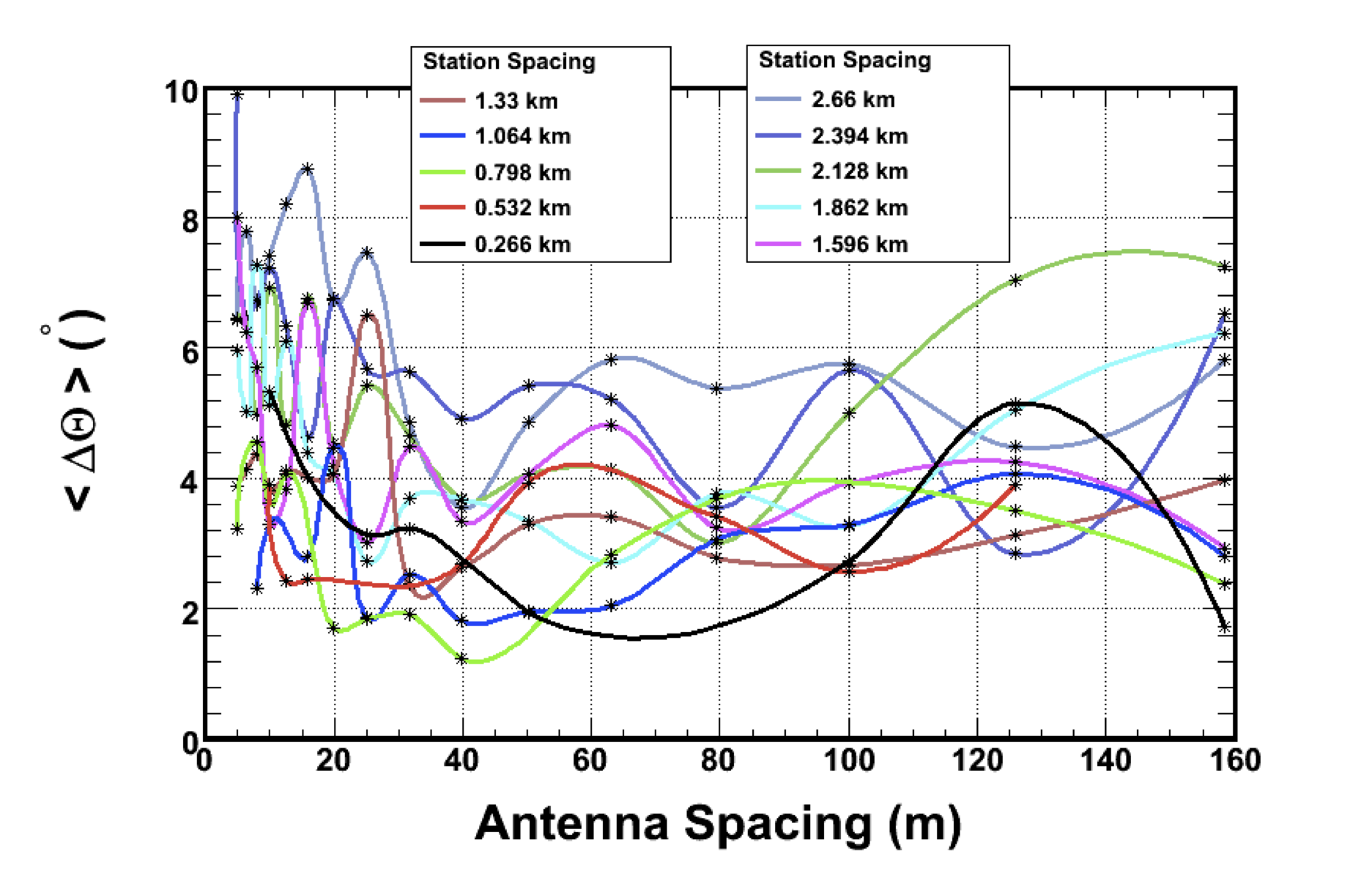}
	\caption{Resolutions of neutrino direction, $\langle\Delta \Theta_\nu\rangle$ versus antenna spacings and station spacings.}
	\label{fig:profile}
	\end{center}
\end{figure}

\begin{figure}
	\begin{center}
	\includegraphics[width=8cm]{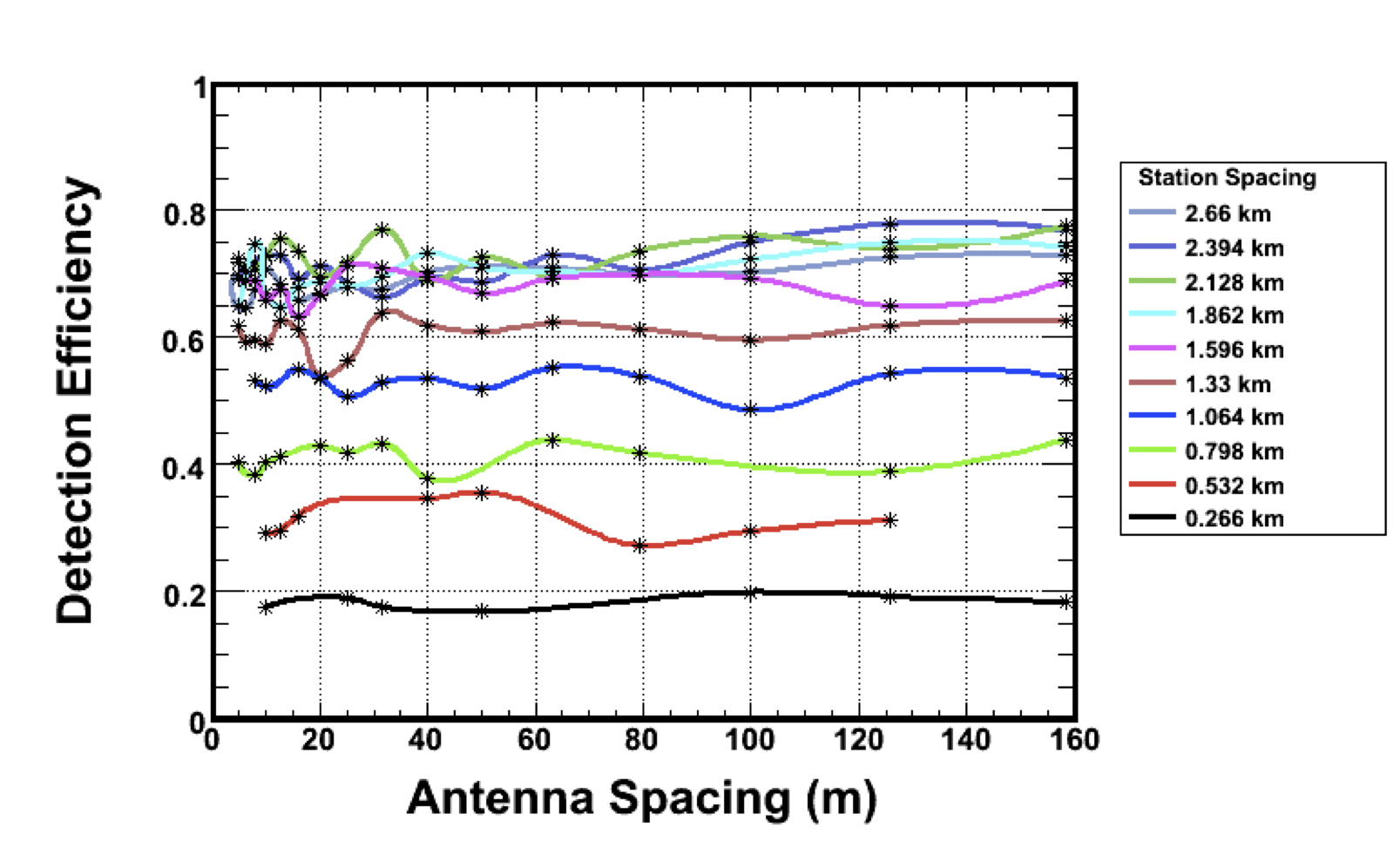}
	\caption{Detection efficiencies versus antenna spacings and station spacings.}
	\label{fig:profile2}
	\end{center}
\end{figure}

     The mean value of the separation angles $\langle\Delta \Theta_\nu\rangle$ versus the antenna spacings is shown in Fig. \ref{fig:profile}, whereas the detection efficiencies versus the antenna spacings are given in Fig. \ref{fig:profile2}.     

%XXXXXXXXXXXXXXXXXXXXXXXXXXXXXXXXXXXXXXXXXXXXXXXXXXXXXX

     Figs. \ref{fig:profile}, and \ref{fig:profile2} suggest that $\langle\Delta \Theta_\nu\rangle$ can be less than 5$^\circ$ if the station spacing is set in the range of 1.33 km to 1.9 km and the antenna spacing is set in the range of 40 m to 100 m. One may notice that the detection efficiency reach a saturated value, $\sim 70$\%, when the station spacing is grater than $\sim 1.5$km.

     To finalize the optimal choice for the ARA geometry, the effects of different noise levels added to the original waveform and different trigger thresholds are studied, too. The value of $\sigma_{noise}$ is set at 0.035 mV for all analysis presented so far with $V_0^{gen}$ varying in the range of 0 to 5 V. In the following studies of how the noise levels would affect the resolution of the neutrino moving direction, in each case a different level of noise added to the waveform is assumed, i.e. $\sigma_{noise}^\prime=\alpha \sigma_{noise}$, with $\alpha$ greater than one, whereas $V_0^{gen}$ is fixed at 5 V. Different trigger thresholds are applied: $V_i^{obs}>3.5\sigma_{noise}$, $V_i^{obs}>7\sigma_{noise}$. For these studies, only 100 events are generated in each case. The results of $\langle\Delta \Theta_\nu\rangle$ and the detection efficiency versus the noise level under different trigger thresholds are presented in Figs. \ref{fig:spa5} to \ref{fig:Nspc7} for different antenna spacings and different station spacings. It was found that the larger $\sigma_{noise}^\prime$ added to the waveforms, the worse the resolution of the neutrino moving direction, which is as expected. In addition, the higher the trigger threshold, the lower the detection efficiency.

\begin{figure}
	\begin{center}
	\includegraphics[width=8cm]{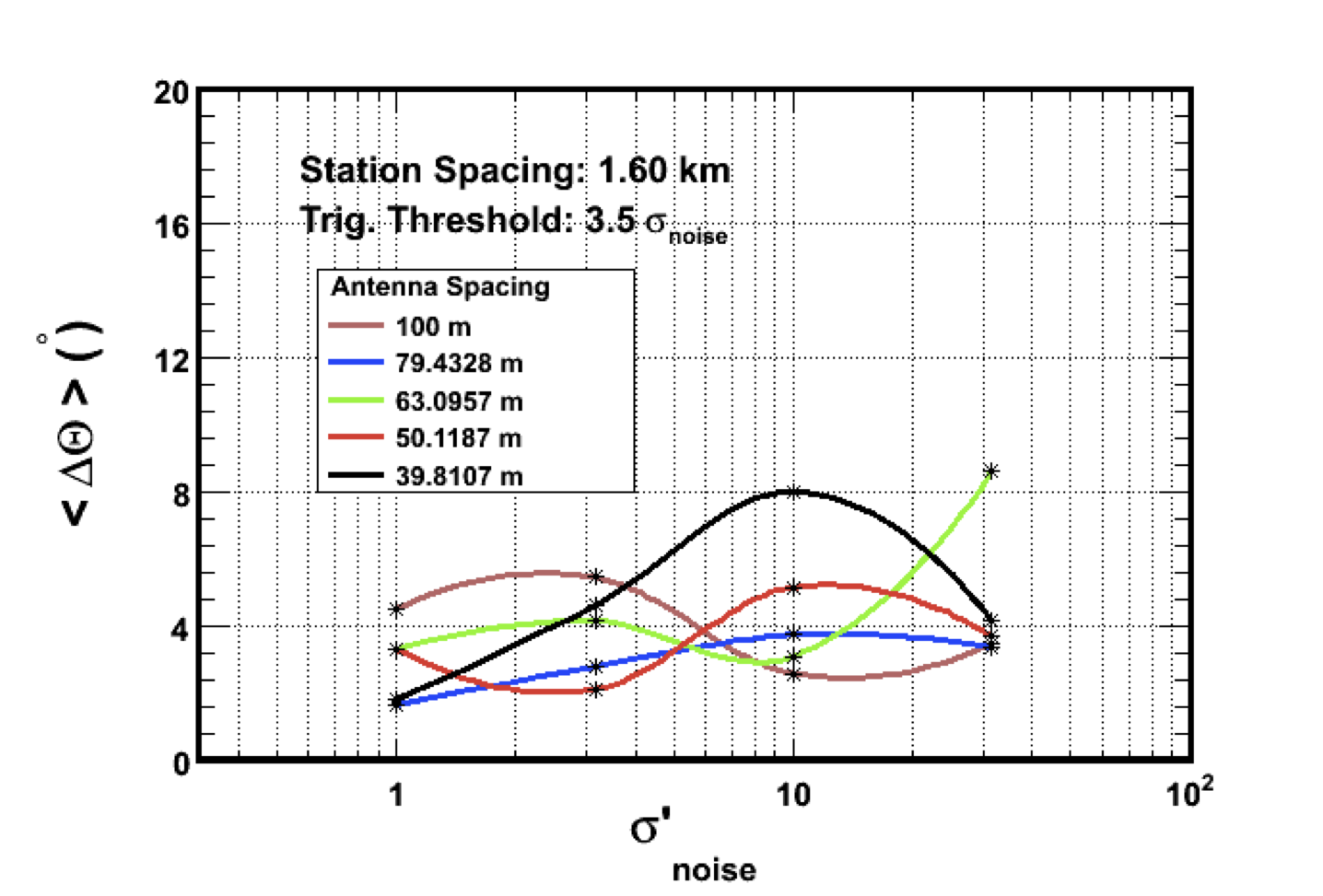}
	\caption{Resolutions of neutrino direction, $\langle\Delta \Theta_\nu\rangle$, versus different noise levels, and antenna spacings, where station spacing is set at 1.60 km and the trigger threshold is $3.5\sigma_{noise}$.}
	\label{fig:spa6}
	\end{center}
\end{figure}

\begin{figure}
	\begin{center}
	\includegraphics[width=8cm]{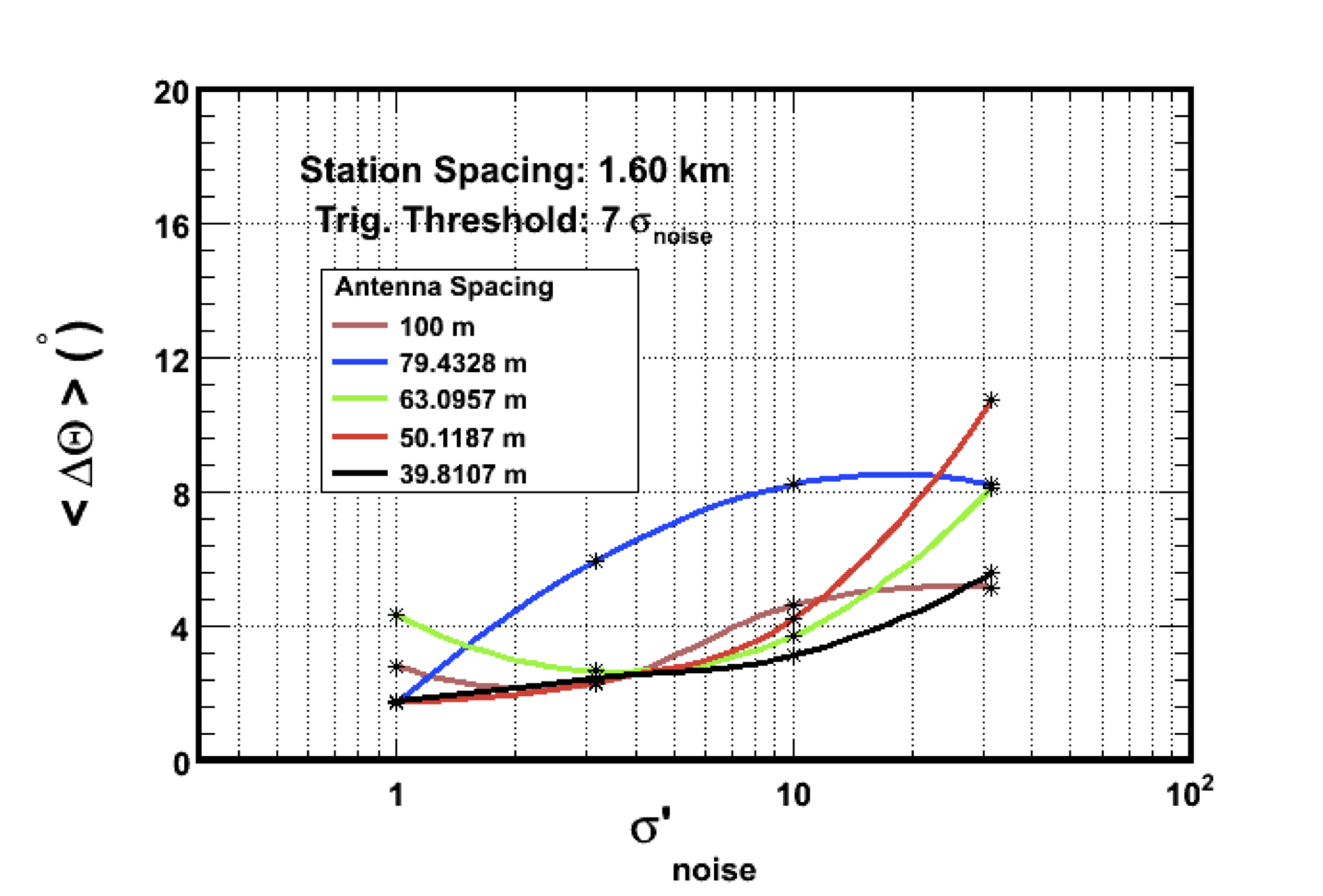}
	\caption{Resolutions of neutrino direction, $\langle\Delta \Theta_\nu\rangle$, versus different noise levels, and antenna spacings, where station spacing is set at 1.60 km and the trigger threshold is $7\sigma_{noise}$.}
	\label{fig:spc6}
	\end{center}
\end{figure}

\begin{figure}
	\begin{center}
	\includegraphics[width=8cm]{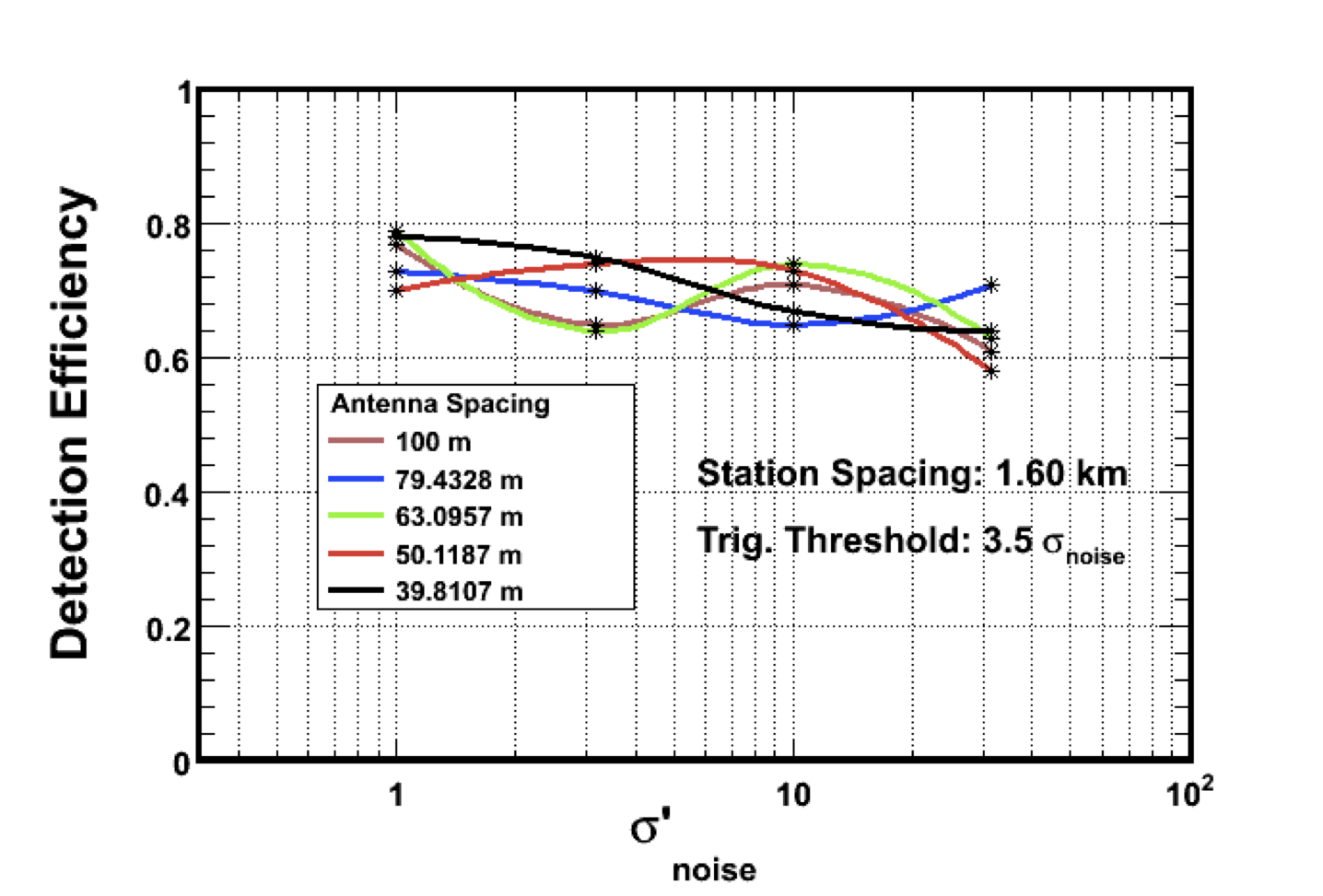}
	\caption{Detection Efficiencies versus different noise levels and antenna spacings,  where station spacing is set at 1.60 km and the trigger threshold is $3.5\sigma_{noise}$.}
	\label{fig:Nspa6}
	\end{center}
\end{figure}

\begin{figure}
	\begin{center}
	\includegraphics[width=8cm]{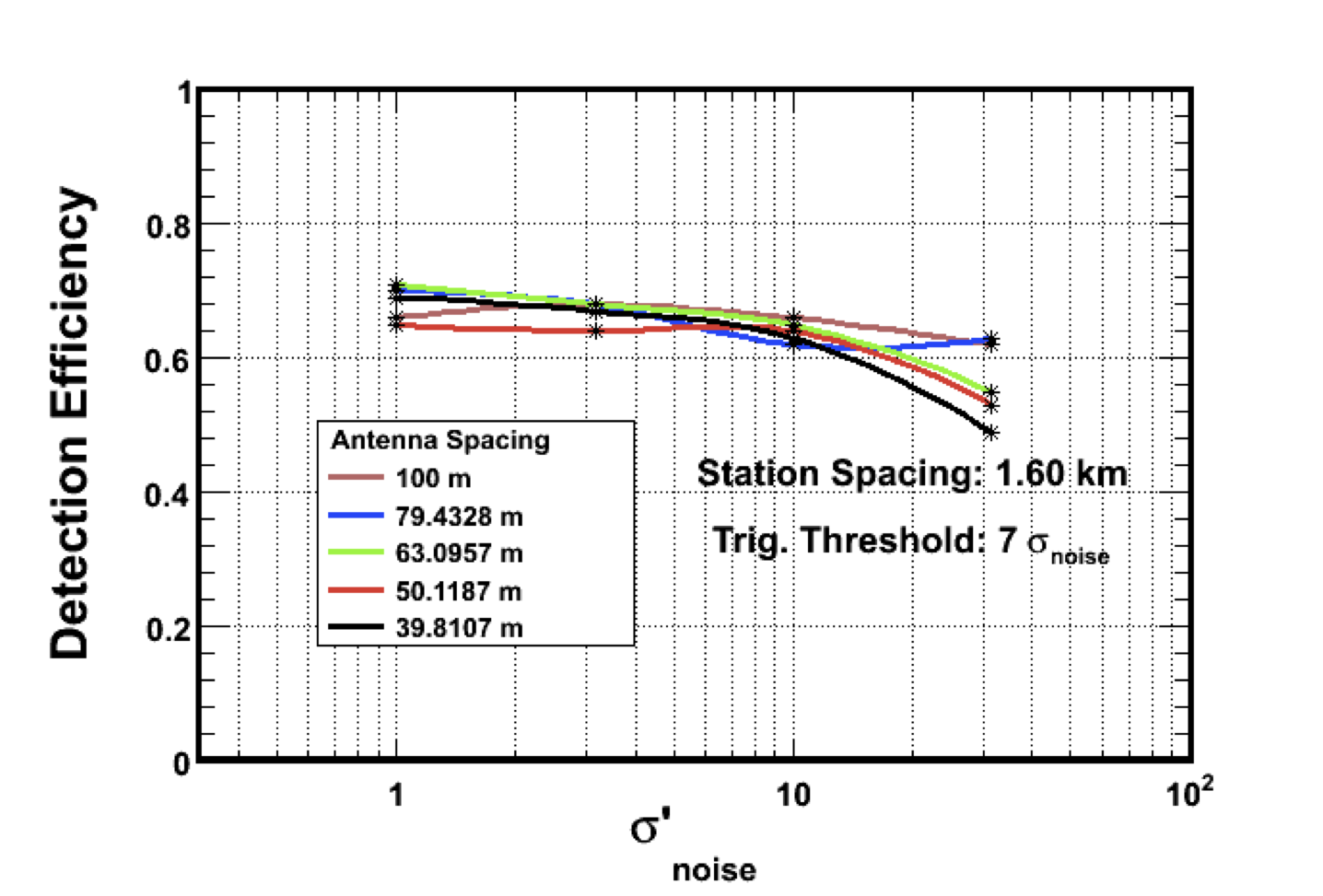}
	\caption{Detection Efficiencies versus different noise levels and antenna spacings,  where station spacing is set at 1.60 km and the trigger threshold is $7\sigma_{noise}$.}
	\label{fig:Nspc6}
	\end{center}
\end{figure}

     In summary, with the noise effect taken into account, in order to make the resolution of the neutrino moving direction as good as possible and detection efficiency as high as possible, the optimal choice for ARA geometry would be 1.6 km for the station spacing and 40 m for the antenna spacing.

%\section{Discussion}
\section{Summary}

     Angular Resolution of Neutrino Moving Direction: One of the main goals of ARA is to point back to cosmic accelerators through the determination of the UHE neutrino moving directions, so the resolution of it is particularly important.

     To optimize the ARA, both the resolution of the neutrino moving direction and the detection efficiency should be considered. Basically, the detection efficiency increases as the station spacing gets larger. From Fig. \ref{fig:profile2}, however, it reaches a plateau of $\sim 70$\% detection efficiency when the station spacing is grater than $\sim$1.5 km where the regions which each station can cover no longer overlap. With the noise effect taken into account, in order to make the resolution of the neutrino moving direction as good as possible and detection efficiency as high as possible, the optimal choice for ARA geometry would be 1.6 km for the station spacing and 40 m for the antenna spacing.  

     In the simulation of angular resolution of neutrino direction for Antarctic Ross Ice Shelf ANtenna Neutrino Array (ARIANNA) experiment, the resolution in $\theta$ direction is $1.1^\circ$ \cite{arianna}. However, to reach such a good resolution, ARIANNA has to build its array up to 11 stations per $\rm{km}^2$, which means that its antenna density has to be 13 times greater than ARA if we set the station spacing as 1.33 km. Based on this comparing, the design of ARA is in a better balance point between the resolution and the cost.

     In the future, if ARA can get more funding to increase the density of the antenna number, a much better resolution of neutrino moving direction can be achieved.

%\section{Acknowledgement}
\section*{Acknowledgement}

We thank the High Performance Grid Computing Center of National Taiwan University and the National Center for High-Performance Computing for their support of computing facility and time that helped expedite our calculations. This work is supported by the National Research Council under Project No. NSC 98-2811-M-002-501. PC is in addition supported by the US
Department of Energy under Contract No. DE-AC03-76SF00515.

\end{document}